\documentclass[prb,aps,twocolumn]{revtex4}
\usepackage{graphicx}
\usepackage{dcolumn}
\usepackage{bm}

\begin{document}

\title{Reply to the Comment on "Hole digging in ensembles of tunneling
molecular magnets"}

\author{I. S. Tupitsyn$^{1,2}$ P.C.E. Stamp$^2$ and N.V. Prokof'ev$^3$}
\affiliation{$^{1}$ Russian Research Center "Kurchatov Institute",
Kurchatov Sq.1, Moscow 123182, Russia. \\
$^2$ Department of Physics and Astronomy, and Pacific Institute
for Theoretical Physics, University of
British Columbia, 6224 Agricultural Rd., Vancouver, B.C. V6T 1Z1, Canada \\
$^3$ Department of Physics, University of Massachusetts, \\
Amherst, MA 01003, USA.}


\begin{abstract}

Our work has argued for a particular scaling form governing the
distribution $M(\xi,t)$ of magnetisation over bias $\xi$, for a
system of dipolar-interacting molecular spins. This form, which
was found in Monte Carlo (MC) simulations, leads inevitably to a
short-time form $\sim t^{1/2}$ for the magnetisation relaxation in
the system. The authors of the Comment argue that the
magnetisation should decay rather as $\sim t^p$, with the exponent
$p$ depending on the lattice type- and they argue this form is
valid up to infinite times. They also claim that our conclusion is
based on an assumed exponential dependence of the function
$M(\xi,t)$ on $\tau_{de}(\xi)$, the effective molecular relaxation
time. In fact our results do not depend on any such dependence,
which was used merely for illustrative purposes, but only on the
scaling form we found. Repeating our MC simulations for
different lattice types and different parameters, we always find a
square root relaxation for short times. We find that the results
of the comment are flawed because they try to fit their results
over far too large a range of times (including the infinite time
limit, where no simple theory applies).

\end{abstract}

\maketitle

The comment \cite{AF04} misrepresents both our work and the
physics of the problem. Our results are essentially that for both
strongly polarised \cite{PS98} and depolarised \cite{TSP04}
dipolar systems, the short-time relaxation (after an initial
transient) has the form $\delta M \sim \sqrt{t}$, regardless of
lattice type- with concomitant results for the hole width and the
scaling function $M(\xi,t)$ introduced in \cite{TSP04}. It is
argued in the comment \cite{AF04} that (i) our results are based
on the assumption of an exponential form for the function
$M(\xi,t)$; and (ii) that the magnetisation relaxation in lattices
of various symmetry may be described over huge time ranges,
encompassing up to 3 orders of magnitude in magnetisation (from
$M=1$ to $M=10^{-3}$) by a power law form $\delta M(t) \sim t^p$,
with $p$ dependent on lattice symmetry. We respond to these in
turn:

\vspace{2mm}

(i) The effective relaxation rate $\tau^{-1}_{de}(\xi)$ was
introduced in \cite{TSP04}, and extracted and studied using direct
$t$-dependent MC simulations. It describes molecules in large bias
fields $\xi$, and, we claimed, has a Lorentzian-tail dependence
$\tau^{-1}_{de}(\xi) \sim \xi^{-2}$. As before we write the total
normalised magnetisation of the system $M(t) = \int d\xi
M(\xi,t)$, where $M(\xi,t)$ is the normalised distribution of
magnetisation over the local bias fields $\xi$ acting on each
molecular spin; in terms of the probability $P_{\sigma}(\xi,t)$
for a spin to be in a bias field $\xi$ at time $t$, one has
$M(\xi,t) = [P_{\uparrow}(\xi,t) - P_{\downarrow}(\xi,t)]$. In
commenting on our paper Alonso and Fernandez have introduced a
function $f(\xi,t)$ which is nothing but $f(\xi,t) = - M(\xi,t)$.
They argue that our conclusions are based on the assumption that
our $M(\xi,t)$ depends on $\tau_{de}(\xi)$ via an exponential
form, ie., that $M(\xi,t) \propto e^{-t/\tau_{de}(\xi)}$. In fact
this is not correct- the crucial conclusion in our paper about
$\tau_{de}(\xi)$ was that it scales as $\xi^2$, so that we can
write $M(\xi,t)$ as a function of one scaled variable alone,
satisfying the scaling law
\begin{equation}
M(\xi,t) \approx M(t / \xi^2 ) \equiv M(z), \label{scaling}
\end{equation}
where $z= t/\tau_{de} \sim t/\xi^2 $. The actual form of $M(z)$ is
not so crucial. In our paper we used an {\it exponential} decay
form for $M(z)$ (this is just the standard $\tau$-approximation
result, and so the simplest form to use). This was used for {\it
illustrative purposes only}, and none of our conclusions is based
on the specific form of this function- contrary to the comment's
claim! In fact it is very easy to see that the functional form of
$M(z)$ is {\it irrelevant} for the final conclusion that the
magnetization decay is $\sim t^{1/2}$, which simply and
immediately follows from the use of the scaling form in $ M(t) =
\int d\xi M(\xi,t)$, ie., {\it from the scaling form alone}. This
scaling form was found to be valid in our MC simulations for
different lattice types, provided one assumes that $t > \tau_o$
(ie., assuming initial transients are over) and that $\delta
M/M(t=0) \ll 1$ (ie., assuming that the long-time multispin
correlations have not yet set in). We also assumed that $\xi_o \ll
\xi \ll W_D$ (here $W_D$ is the half-width of the dipole energy
distribution, $\xi_o$ is the nuclear spin bath parameter, and
$\tau_o$ is the characteristic single-molecule relaxation time
\cite{TSP04}- hereafter we measure time in units of $\tau_o$ and
energies in units of $\xi_o$).

\vspace{2mm}

\begin{figure}[tbh]
\centering
\vspace{-2.0cm}
\hspace{0.cm}
\includegraphics[scale=0.4]{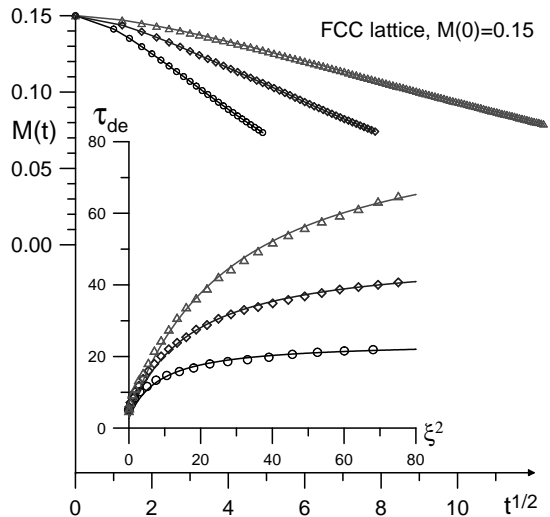}
\vspace{-2.3cm}
\caption{$M(t)$ vs $\sqrt t$ in the FCC lattice
for initial magnetization $M(0)=0.15$ and three values of $E_D$:
$E_D=2.5$  ({\it circles}), $E_D = 5$  ({\it diamonds}) and $E_D =
10$ ({\it triangles}). The inset shows the effective relaxation
time $\tau_{de}(\xi)$ vs $\xi^2$. Solid lines in the inset are the
shifted Lorentzian curves described in the text with $\alpha(E_D)
\approx 0.0415, 0.0205, 0.0114$ for $E_D = 2.5, 5, 10$. The
time-step in the MC simulations was $\delta t / \tau_o =
10^{-2}$.}
\label{fig:tsp_repf}
\end{figure}

(ii) Alonso and Fernandez are basically arguing that the scaled
variable $z$ should be rather written as $t / \xi^{1/p}$ with the
exponent $p$ depending on the lattice type; correspondingly, $M(t)
\sim t^p$ and for FCC lattices they found $p \simeq 0.73$. For SC
lattices they found $p=0.5$. In Fig.~\ref{fig:tsp_repf} we present
MC solutions of kinetic equations for the magnetization decay in
the FCC lattice with $60^3$ spins (FCC, BCC and triclinic lattices
were also analyzed in Ref. \cite{TS04}). After the usual initial
transient behavior $\sim t$ (already discussed in Ref. \cite{PS98}
as well as \cite{TSP04}), we observe a long interval with the
$\sim \sqrt{t}$ relaxation at least until the fractional change in
$|M(t) - M(0)| / M(0) \sim 0.5$. At longer times relaxation
deviates from the $\sim t^{1/2}$ behavior, but the duration of the
$\sim t^{1/2}$ interval in a demagnetized sample is {\it longer}
than in a polarized sample (compare figures in Refs. \cite{PS98}
and \cite{TSP04,TS04}).

Simultaneously, we find that the functional form of the effective
relaxation rate in the FCC lattice is well described by the
shifted Lorentzian $\tau^{-1}_{de}(\xi) = (1 / \pi) / (\xi^2 + 1)
+ \alpha(E_D)$ with small $\alpha(E_D) \sim 1/W_D$ (again, $\xi_o
= 1$ in our units); at small values of initial polarization in the
FCC lattice $W_D \approx 10 E_D$, where $E_D$ is the strength of
the dipole-dipole interactions between nearest neighbor spins. We
did not find any qualitative difference between the SC, triclinic,
BCC and FCC cases.

It is important to notice that when the number of molecules in
resonance is very small (as it is in the limit of very small
values of $\xi_o / W_D$, particularly for FCC lattices), the usual
{\it initial transient} \cite{PS98} of $M(t)$ can be rather long,
and its duration sensitive to the crystal structure because the
dipole field spectrum is discrete in this limit. Then MC results
on a small finite system are not relevant to experiments on
macroscopic samples (where the number of molecules in resonance
is also macroscopic).

In the comment, and in a previous paper \cite{FAPRB04}, the
authors have argued that they can fit the whole time range,
including both very short and even {\it infinitely} long times,
using a single theory with a single power law exponent $p$ (see,
eg., Figs. 3,4 of the comment and Ref. \cite{FAPRB04}, paragraphs
before Eqtn. 22 and after Eqtn. 24 of this paper). We find this
implausible. Neither our MC simulations \cite{PS98,TSP04,TS04,PS98a}
nor our analytic work \cite{PS98} have ever claimed or attempted to
explain more than the initial transient and the short-time relaxation
that ensues after this transient, in the time interval before
multi-spin correlations build up. It is in this restricted time
range that we have argued for simple scaling and the associated
square root relaxation.


\begin{thebibliography}{99}

\bibitem{AF04} J.J. Alonso, J.F. Fernandez, preceding comment

\bibitem{PS98} N.V. Prokof'ev and P.C.E. Stamp, Phys. Rev. Lett.,
{\bf 80}, 5794 (1998).

\bibitem{TSP04} I.S. Tupitsyn, P.C.E. Stamp and N.V. Prokof'ev.
Phys. Rev. B {\bf 69}, 132406, (2004).


\bibitem{TS04} I.S. Tupitsyn and P.C.E. Stamp, Phys. Rev. Lett.
{\bf 92}, 119701 (2004).

\bibitem{PS98a} N.V. Prokof'ev and P.C.E. Stamp, J Low Temp. Phys.
{\bf 113}, 1147 (1998)

\bibitem{FAPRB04} J.F. Fernandez and J.J. Alonso, Phys. Rev. B
{\bf 69}, 024411 (2004). Analytical calculations of the exponent
$p$ were claimed in this paper, which assumed the form $\delta M
\sim t^p$ to be valid both up to the limit $t \to \infty$, and
also in the limit of vanishingly small values of $\xi_o / W_D$. No
account was taken of either the discreteness of the dipole field
spectrum in the small $\xi_o$ limit, or of higher multi-spin
correlations in the long time limit.

\end{thebibliography}
\end{document}